\documentclass[
 reprint,
 floatfix,
 amsmath,amssymb,
 showpacs,
 superscriptaddress,
 citeautoscript,
 nonatbib
]{revtex4-1}

\usepackage{xr}
\makeatletter
\newcommand*{\addFileDependency}[1]{
  \typeout{(#1)}
  \@addtofilelist{#1}
  \IfFileExists{#1}{}{\typeout{No file #1.}}
}
\makeatother

\usepackage{amssymb}

\usepackage[breaklinks]{hyperref}
\usepackage[all]{hypcap}

\hypersetup{
    plainpages=false,
    unicode=false,                          
    pdfborder={0 0 0},
    pdftoolbar=true,                        
    pdfmenubar=true,                        
    pdffitwindow=false,                     
    pdfstartview={FitH},                    
    pdftitle={},                            
    pdfauthor={},                           
    pdfproducer={LNCMI-CNRS-UGA-UPS-INSA},  
    pdfkeywords={High} {Magnetic} {Fields}, 
    pdfnewwindow=true,                      
    linktoc=section,
    colorlinks=true,                        
    linkcolor=blue,                         
    citecolor=red,                          
    filecolor=magenta,                      
    urlcolor=blue                           
}

\usepackage{graphicx}
\usepackage{float}
\usepackage{amsmath}
\usepackage{bm}
\usepackage{color, soul}
\usepackage[normalem]{ulem}
\usepackage{siunitx}
\usepackage{upgreek}
\usepackage[usenames,dvipsnames]{xcolor}
\usepackage{textcomp}
\usepackage{chemformula}
\usepackage[normalem]{ulem}

\usepackage{booktabs}
\usepackage{amssymb}
\usepackage{colortbl}   

\makeatletter
\AtBeginDocument{\let\hl\@firstofone}
\makeatother

\begin{document}

\author{Stefan A. Seidl}
\thanks{These authors contributed equally.}
\affiliation{Physics Department, TUM School of Natural Sciences, Technical University of Munich, 85748 Garching, Germany\looseness=-1}
\author{Xiangzhou Zhu}
\thanks{These authors contributed equally.}
\affiliation{Physics Department, TUM School of Natural Sciences, Technical University of Munich, 85748 Garching, Germany\looseness=-1}
\author{Guy Reuveni}
\affiliation{Department of Chemical and Biological Physics, Weizmann Institute of Science, Rehovot 76100, Israel\looseness=-1}
\author{Sigalit Aharon}
\affiliation{Department of Chemical and Biological Physics, Weizmann Institute of Science, Rehovot 76100, Israel\looseness=-1}
\author{Christian Gehrmann}
\affiliation{Physics Department, TUM School of Natural Sciences, Technical University of Munich, 85748 Garching, Germany\looseness=-1}
\author{Sebasti\'{a}n Caicedo-D\'{a}vila}
\affiliation{Physics Department, TUM School of Natural Sciences, Technical University of Munich, 85748 Garching, Germany\looseness=-1}
\author{Omer Yaffe}
\affiliation{Department of Chemical and Biological Physics, Weizmann Institute of Science, Rehovot 76100, Israel\looseness=-1}
\author{David A. Egger}
\email{david.egger@tum.de}
  \affiliation{Physics Department, TUM School of Natural Sciences, Technical University of Munich, 85748 Garching, Germany\looseness=-1}

\title{Anharmonic Fluctuations Govern the Band Gap of Halide Perovskites}

\date{\today}

\begin{abstract}
\noindent We determine the impact of anharmonic thermal vibrations on the fundamental band gap of \ch{CsPbBr3}, a prototypical model system for the broader class of halide perovskite semiconductors.
Through first-principles molecular dynamics and stochastic calculations, we find that anharmonic fluctuations are a key effect in the electronic structure of these materials.
We present experimental and theoretical evidence that important characteristics, such as a mildly changing band-gap value across a temperature range that includes phase-transitions, cannot be explained by harmonic phonons thermally perturbing an average crystal structure and symmetry.
Instead, the thermal characteristics of the electronic structure are microscopically connected to anharmonic vibrational contributions to the band gap that reach a fairly large magnitude of 450\,meV at 425\,K.
\end{abstract}
\maketitle
\noindent The fundamental band gap of a semiconductor in the single-particle picture is defined as the energy difference between the valence band maximum and conduction band minimum.
Determining the color of crystalline materials, it characterizes an important energy scale for light-matter interactions, relevant for semiconductor applications in optoelectronic devices.
For example, in photovoltaics the maximum reachable efficiency of a single-junction solar cell is dictated by the band gap of the absorber material \cite{wurfel2009,shockley1961}.
Because technological devices are used around room temperature, it is necessary to rationalize and predict the impact of thermal effects on the band gap and other optoelectronic properties of semiconducting materials.\\
There is an ongoing debate on the leading thermal effects that determine band gaps of halide perovskites (HaPs), \hl{which are} promising semiconductors for optoelectronics.
Specifically, higher-order electron-phonon interactions \cite{saidi2016a,saidi2018a}, thermal lattice expansion \cite{yu2011,huang2013,francisco-lopez2019}, as well as  vibrations~\cite{patrick2015b,whalley2016,yang2017a,lanigan-atkins2021} were shown to influence the band gap.
Interestingly, several theoretical studies found that an \textit{average band-gap value}, computed for configurational samples of HaPs in quasi-static~\cite{yang2017a,zhao2020,zacharias2023} or fully dynamical calculations \cite{carignano2015,quarti2016,wiktor2017,lanigan-atkins2021}, can deviate significantly from a band-gap value that is obtained in \textit{static calculations on the average crystal structure}.
Therefore, it is currently discussed that optoelectronic properties of HaP semiconductors are impacted by their rather unusual dynamic structural responses at elevated temperatures.
These include pronounced octahedral tilting dynamics and large-amplitude atomic displacements, which both are microscopically connected \textit{via} the effect of vibrational anharmonicity \cite{lanigan-atkins2021,weadock2023a,beecher2016,yang2017a,yaffe2017,wu2017,gold-parker2018a, klarbring2019a,liu2019,ferreira2020,debnath2021,fransson2022b,zhu2022a,cannelli2022b}.\\
\hl{Vibrational anharmonicity describes atomic motions which go beyond the independent phonon picture that is rooted in the harmonic approximation.
It assumes that finite-temperature displacements of atoms in a periodic arrangement change the potential energy of the system merely quadratically.
Anharmonic fluctuations become relevant when the actual atomic displacements sample parts of the potential energy surface deviating from this quadratic form.}
\textcolor{Black}{HaPs exhibit strong anharmonicity to the extent that their vibrations are localized in real space~\cite{yaffe2015, ferreira2020, weadock2023a}.}\\
Indeed, \hl{for HaPs} impacts of such dynamic fluctuations on various quantities related to the band gap have been reported, \textit{e.g.}, for defect characteristics \cite{cohen2019,wang2019,kumar2020,zhang2020,chu2020,park2022}, carrier mobilities \cite{mayers2018a,lacroix2020,schilcher2021,lai2022,iaru2021a,zhang2022a}, exciton properties \cite{filip2021}, and Urbach energies \cite{wu2019,gehrmann2019,gehrmann2022}.
However, the precise connections between anharmonic structural fluctuations and thermal effects in the band gap have not yet been characterized for HaP semiconductors.
Resolving how anharmonic structural fluctuations impact band gaps of semiconductors is an opportunity to quantify emergent thermal effects in the electronic structure, which is a challenge for both experimental and theoretical methods.
For the class of HaP materials, knowledge on how anharmonic structural fluctuations impact the band gap will deliver microscopic understanding relevant for their technological application and future material design.\\
In this letter, we quantify the effect of anharmonic structural fluctuations on the electronic structure of HaPs by studying the temperature evolution of the band gap in a prototypical HaP variant, \ch{CsPbBr3}, across phase transitions.
Experimental data shows that the phase transitions have a minor effect on the band gap, in contrast to predictions of static density functional theory (DFT) calculations, where the high-temperature phase has a significantly smaller band gap than the low-temperature phase.
We then investigate the temperature evolution of the band gap through first-principles calculations, using stochastic Monte-Carlo (MC) and molecular dynamics (MD) techniques, where the former includes only harmonic structural fluctuations while MD includes all atomic motions.
Comparing the results of both computations in operationally relevant temperature regimes, we establish that anharmonic fluctuations are a key effect for the electronic structure of these systems, reaching a comparatively large magnitude of 450\,meV for the band gap in the cubic phase of \ch{CsPbBr3}.\\
\begin{figure}[!]
	\centering
	\includegraphics[width=1.\columnwidth]{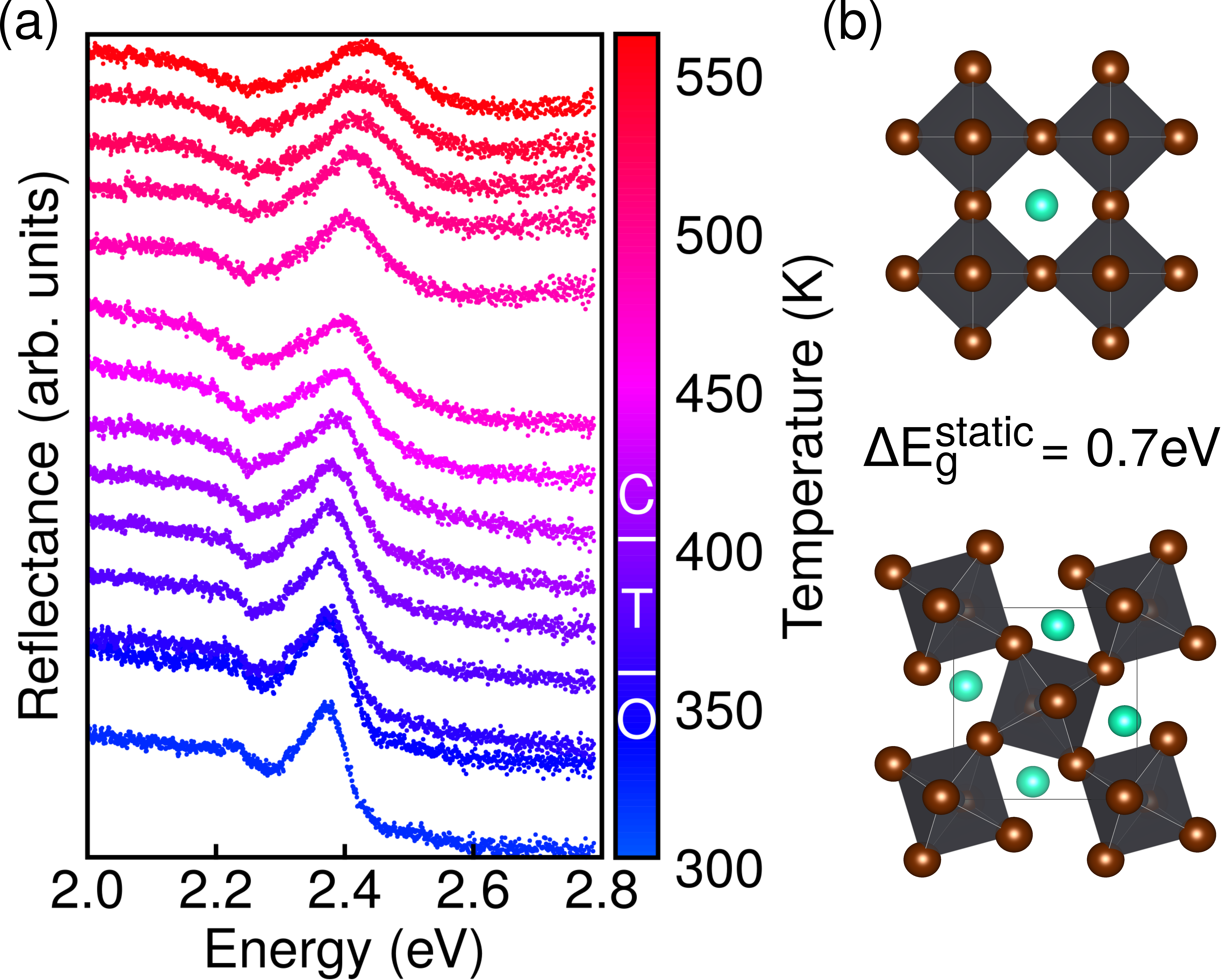}
	\caption{(a) Temperature-dependent reflectance measurements of \ch{CsPbBr3} single crystals. The color-bar on the right defines temperature in the spectra, which was measured at 20~K intervals. Thin horizontal lines indicate reported phase-transition temperatures \cite{hirotsu1974}.
	Schematic representations and band-gap change, $\Delta \mathrm{E}_\text{g}^\text{static}$, computed in static DFT calculations of the orthorhombic (panel b) and cubic structure (panel c) of \ch{CsPbBr3}.
	}
	\label{fig1}
\end{figure}%
Fig.~\ref{fig1}\textcolor{blue}{a} (see \textcolor{Black}{Appendix}~\ref{app:b} for further details) presents the temperature evolution (300$-$560\,K) of the reflectance spectra of a \ch{CsPbBr3} single-crystal.
A peak at ${\approx}2.4$~eV, associated with excitonic absorption \cite{Fox2010}, dominates the spectra.
Since the exciton binding-energy of \ch{CsPbBr3} is on the order of a few tens of meV's \cite{yettapu2016}, the temperature evolution of the peak is a proxy for the temperature evolution of the band gap.
Notably, thermally-induced changes in the spectra are mild despite the orthorhombic-to-tetragonal and tetragonal-to-cubic phase transitions at 361~K and 403~K, respectively \cite{hirotsu1974,stoumpos2013}.
These findings agree with previous experimental work~\cite{mannino2020}, but are in stark contrast with results of static theoretical calculations, showing a significant decrease of up to 0.7\,eV in the band-gap value between the orthorhombic and cubic structure \cite{murtaza2011,kang2018}.
To emphasize this point, Fig.~\ref{fig1}\textcolor{blue}{b} shows results from static DFT calculations, where we employ the PBE functional \cite{perdew1996a}, augmented by dispersive corrections \cite{tkatchenko2009,bucko2014}, and accounting for spin-orbit coupling (SOC) as implemented in \texttt{VASP} \cite{kresse1996b} (see \textcolor{Black}{Appendix}~\ref{app:a} for further details).
This experiment-theory discrepancy motivates in-depth theoretical inspection of leading effects in the finite-temperature electronic structure of HaPs. \\
We begin by investigating the effect of temperature within the harmonic approximation, \textit{i.e.}, phonon-induced gap renormalization.
Fig.~\ref{fig2} shows theoretical data obtained in a DFT-MC method \cite{karsai2018,zacharias2015,monserrat2014c}, which includes higher-order terms in the perturbation of the electronic structure due to temperature-activated phonons.
\hl{The MC method provides thermal samples of atomic displacements according to a pre-calculated phonon spectrum (see \textcolor{Black}{Appendix}}~\ref{app:a}\hl{ for details).
It thereby strictly enforces the harmonic approximation and assumes an average crystal structure as a starting point for the finite-temperature treatment of the electronic states.}
Using the PBE exchange-correlation functional and accounting for SOC as above, we adopt it for the cubic and orthorhombic phase, omitting the intermediate tetragonal structure that only exists in a very narrow temperature range.\\
The MC method yields band-gap values of \ch{CsPbBr3} at 325\,K and 425\,K that differ by 0.4\,eV.
These results are an improvement compared to the static calculations shown in Fig.~\ref{fig1}\textcolor{blue}{b}, but are still far from the experimental observation showing a mild shift (Fig.~\ref{fig1}\textcolor{blue}{a}).
Consequently, the major discrepancy of Fig.~\ref{fig1} remains largely unresolved even after the thermal effect of harmonic phonons on the band gap was taken into account.\\
\begin{figure}[b!]
	\centering
	\includegraphics[width=1.\columnwidth]{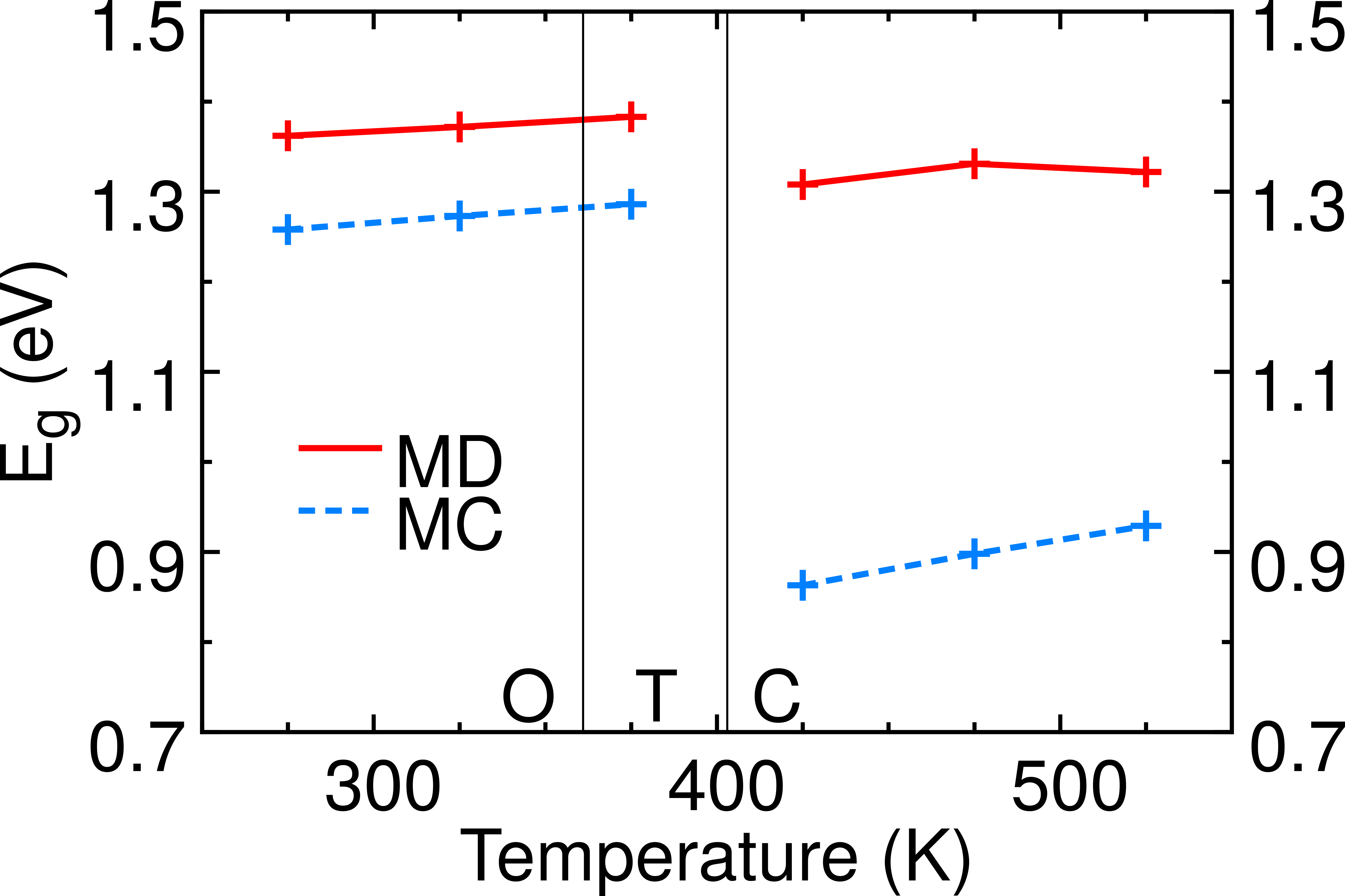}
	\caption{Temperature-dependent fundamental band-gap of \ch{CsPbBr3}, computed in DFT-based MC (blue-dashed lines) and MD calculations (red lines).
	Thin vertical lines indicate reported phase-transition temperatures \cite{hirotsu1974}.}
	\label{fig2}
\end{figure}%
Next, we performed DFT-based MD calculations of the band gap in order to account for vibrational anharmonicity, at various temperatures including those of the orthorhombic and cubic phases of \ch{CsPbBr3}, again employing the PBE DFT-functional.
\hl{MD does not rely on the harmonic approximation and is an exact theory for semiclassical treatments of atomic motions at finite temperature, \textit{i.e.}, it accounts for vibrational anharmonicity to all orders.}
We complete equilibration ($\sim$8\,ps) and extensive production runs ($\sim$42\,ps) to sample the dynamic distortions at a given temperature (see \textcolor{Black}{Appendix}~\ref{app:a}).
Subsequently, we conduct electronic-structure calculations including SOC, on 100 MD-snapshots out of the first $\sim$16\,ps of the production run, to calculate average band gaps.
Combining the MC and MD data allows for extracting the precise amount by which anharmonicity impacts the electronic structure.
As a consistency check, we show it is negligible for bulk Si that generally is expected to be fairly harmonic \cite{SM}.
\\
\begin{figure}[!]
	\centering
	\includegraphics[width=1.\columnwidth]{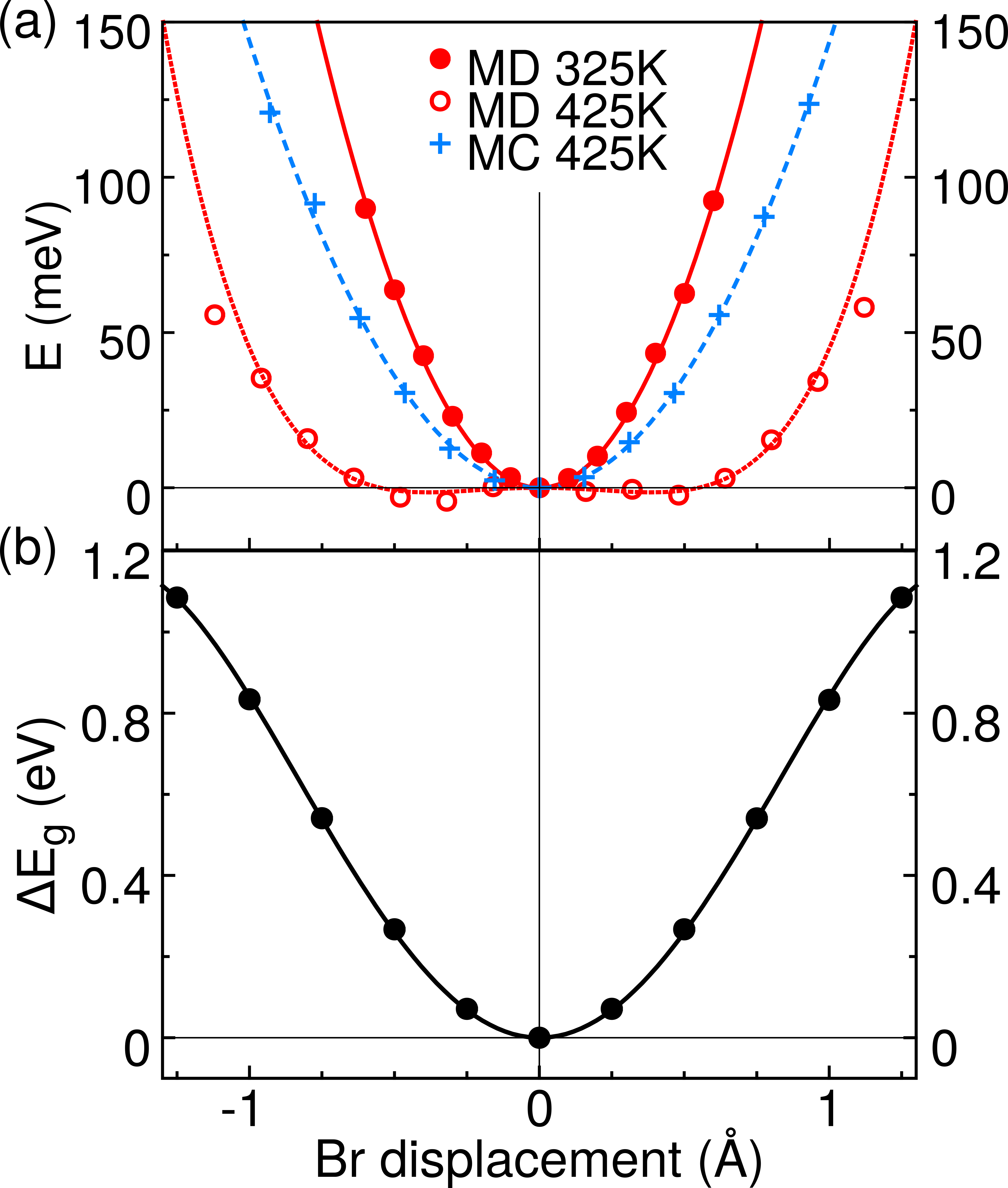}
	\caption{(a) Free-energy changes associated with transversal motion of a Br atom (perpendicular to Pb-Br-Pb bond axis) in \ch{CsPbBr3}, as obtained from Boltzmann-inversion of MD trajectories in the orthorhombic (T=325\,K, red disks) and cubic phase (T=425\,K, red circles).
	\hl{Equivalent data are shown for MC-generated structures of the cubic phase (T=425\,K, blue crosses).
	Colored lines represent fits: the orthorhombic MD and cubic MC data can be fit by a quadratic function, the cubic MD data by a function of the form: $a x^4+bx^2$.}
	Zero on the $x$-axis corresponds to the time-averaged position of a Br atom.
	\hl{(b) DFT-calculated change of the band gap, $\Delta E_\mathrm{g}$, as a function of transversal Br displacement for a set of structures generated using the phonon eigenvector of an octahedral-tilting phonon mode (see Ref.}~\cite{SM} \hl{for details).}
	}
	\label{fig3}
\end{figure}%
Notably, the MD data (see Fig.~\ref{fig2}) capture various key characteristics of the thermal electronic structure in \ch{CsPbBr3}:
first, in the temperature-range between {275\,K and 525\,K}, we find mild changes in the band-gap value, \textit{e.g.}, it differs by ${\approx}60$\,meV in the orthorhombic phase at 325\,K and cubic phase at 425\,K.
Second, taking into account dynamic fluctuations beyond harmonic motions in MD strongly opens up the band gap by ${\approx}0.5$\,eV in the cubic phase at 425\,K compared to the MC result, while the difference amounts to only 0.1\,eV in the orthorhombic case at 325\,K.\\
Therefore, we find that the here extracted \textit{anharmonic effect for the gap} exceeds the pure-phonon contribution and largely resolves the discrepancies of Fig.~\ref{fig1}.
The dominance of the anharmonic effect on the band gap can be further established through a comparison of its magnitude (${\approx}0.5$\,eV) to significant effects of SOC (${\approx}0.8$\,eV) and approximate exchange-correlation treatments (${\approx}0.9$\,eV) that are commonly known for \ch{CsPbBr3} and other HaP variants \cite{SM}.
Consequently, in the cubic phase at 425\,K, the MD-calculated gap of 1.3\,eV is \hl{closer} to the experimentally-measured peak of optical reflectance  (${\approx}2.4$\,eV) than the MC result.
\hl{Remaining discrepancies are related to established deficiencies of the PBE exchange-correlation functional:
the band gap values are generally too small in our calculations because of the well-known band-gap problem in semilocal DFT.}
Ref.~\citenum{SM} \hl{discusses this along with further, potentially important effects for the absolute value and temperature dependence of the gap, \textit{e.g.}, excitonic effects, thermal expansion and volume fluctuations.}\\
To gain insight on the atomic motions governing the temperature evolution of the band gap, we analyze the free-energy changes associated with transversal Br displacements (see Fig.~\ref{fig3}\textcolor{blue}{a}), using
Boltzmann inversion of the MD trajectories in the orthorhombic and cubic phase \cite{SM}.
We analyze the Br displacements because previous studies showed that octahedral tilting dynamics, involving transversal motions of Br atoms \cite{lanigan-atkins2021,zhu2022a,gehrmann2022}, strongly affect the electronic structure of HaPs \cite{Knutson2005,Amat2014,carignano2015,quarti2016,wiktor2017,yang2017a,zhang2020,lanigan-atkins2021,Ning2022}.
In the orthorhombic phase, the free energy is close to being harmonic and thermal atomic displacements are relatively small.
However, in the cubic phase they are anharmonic since they follow a very shallow and broad potential that does not fit a quadratic form.
\hl{Note that free-energy changes in the MC approach are necessarily harmonic as confirmed in  Fig.}~\ref{fig3}\textcolor{blue}{a}\hl{ and Ref.}~\cite{SM}.\\
Interestingly, free-energy changes associated with Br motion are different in orthorhombic and cubic \ch{CsPbBr3}, while the average band gap is relatively similar.
This suggests that the system adopts lower-symmetry structures in the cubic phase, as discussed in previous work \cite{carignano2015,quarti2016,wiktor2017,yang2017a,zhang2020,lanigan-atkins2021,Ning2022}.
Ref.~\cite{SM} \hl{contains analysis of the anharmonic fluctuations showing that they feature configurations that are far-off the average cubic structure and symmetry.
Specifically, the system transiently adopts octahedral tiltings that resemble the \textit{average structure} of the orthorhombic phase.}
This explains the mild band-gap changes in \ch{CsPbBr3} regarding temperature and phase transitions because the transient, lower-symmetry distortions appear on timescales of thermal effects in the electronic structure.\\
\hl{Having established the presence of transiently occurring octahedral tiltings in the cubic phase of CsPbBr$_3$, we examine how vibrational anharmonicity impacts the band-gap changes accompanying these tiltings.
\textcolor{Black}{Connecting the dynamic band-gap changes that occur in the MD to phonon modes is in general a difficult task, because purely harmonic modes leave aside any anharmonic effects.
For example, a calculation of only the band-gap changes accompanying a specific phonon mode does not account for interactions of that mode with other vibrations present in the system.}
A simplified model is constructed in order to provide qualitative insight on the anharmonic effect:
the phonon eigenvector of an octahedral tilting mode in cubic CsPbBr$_3$ is extracted \textit{via} harmonic phonon calculations and the \texttt{phonopy} package} \cite{togo2015a}.
\hl{This particular mode is imaginary (see Ref.}~\cite{SM}) \hl{and not included in the standard MC approach.
\textcolor{Black}{It is imaginary in harmonic calculations of the cubic phase because it represents a phonon mode which is involved in distortions of the crystal structure that lower the symmetry of the material~\cite{yang2017a}.}
It provides us with a set of structures with Br atoms displaced according to this phonon mode.
Next, we compute the change of the band gap, $\Delta E_\mathrm{g}(x)$, as a function of transversal Br displacement, $x$, for this particular phonon mode using DFT (see Fig.}~\ref{fig3}\textcolor{blue}{b}).
\hl{A thermal integration of these band-gap changes can now be performed classically, using either the MC or MD free-energy profile (see Fig.}~\ref{fig3}\textcolor{blue}{a}).
\hl{Therefore, we calculate}
\begin{equation}
\Delta E_\mathrm{g}^\mathrm{tot}=Z^{-1}\int\mathrm{d}x\,\,\Delta E_\mathrm{g}(x)\,\mathrm{exp}(-V_{i}(x)/k_\mathrm{B}T),
\end{equation}
\hl{where $Z=\int\mathrm{d}x\,\,\mathrm{exp}(-V_{i}(x)/k_\mathrm{B}T)$, the index $i$ denotes whether the MC or MD free-energy profile of Fig.}~\ref{fig3}\textcolor{blue}{a}\hl{ is used in the integration, and $T=425\,$K.
Note that this assumes that all Br displacements involved in this particular mode follow the same free-energy profile.
From this simplified model, we obtain a total change of the band gap that is $0.17\,$eV for the MC and $0.28\,$eV for the MD case.
Since the MC free-energy profile is harmonic and the one from MD anharmonic, these results show that larger dynamic changes in the band gap will occur because of anharmonic fluctuations.}
\textcolor{Black}{Therefore, the larger atomic displacements as well as the interactions between vibrational modes that are present in the MD of the slow octahedral dynamics are the most relevant anharmonic effects for the band gap.
While the structural details of the anharmonic effect are specific to HaPs, we hypothesize that it plays a role in anharmonic semiconductors more broadly.}
\\
Finally, we discuss the relevance of our findings for other variants of HaPs, which typically exhibit sequences of lower-to-higher symmetry phase transitions too.
Considering \ch{MAPbBr3} as a common example for 3D hybrid organic-inorganic HaPs, it exhibits an orthorhombic-to-tetragonal and tetragonal-to-cubic phase transition at ${\approx}150$~K and ${\approx}240$~K~\cite{poglitsch1897}, respectively.
Experiments found the lower-temperature transition to induce a noticeable but small (${\approx}10$~meV) lowering of the optical transition energy, while the higher-temperature transition is continuous and often even difficult to spot \cite{Wright2016,mannino2020,Tilchin2016,Guo2019}.
By contrast, static DFT calculations predict a 0.6~eV lowering of the band gap between the tetragonal and cubic phase of \ch{MAPbBr3} \cite{Mosconi2016}.
Our finding that anharmonic fluctuations govern the thermal electronic-structure characteristics and result in mild changes of the band gap may resolve this discrepancy:
at lower temperature these fluctuations are less profound, and a somewhat more noticeable shift of the band gap at the orthorhombic-to-tetragonal transition can be expected from the change of the average structure and symmetry of the crystal, \textit{i.e.}, similar to the templating effect discussed in the literature \cite{Knutson2005,Amat2014}.
However, for the tetragonal-to-cubic transition of \ch{MAPbBr3} at higher temperature, the effect of anharmonic fluctuations dominates and the band gap remains continuous, in line with previous theoretical \cite{quarti2016} and experimental work \cite{Sharma2020} on the related \ch{MAPbI3} compound.
Interestingly, in the class of 2D hybrid organic-inorganic HaPs somewhat more profound spectral changes in the optical response have been reported upon a phase transition even at 270~K \cite{Ishihara1990,yaffe2015,Ziegler2022}.
Both static DFT calculations and optical measurements are in agreement that a lowering of the band gap occurs from the lower-temperature to the higher-temperature phase, signaling that the change in the average structure and symmetry is important \cite{Ziegler2022}.
However, again the theoretical shift was found to be significantly larger than the experimental one (by 120~meV) \cite{Ziegler2022}, which in light of our findings may be resolved by anharmonic fluctuations that also occur in 2D HaPs \cite{Menahem2021}.
The interplay of static and dynamic effects in the optoelectronic properties of HaPs motivates future research on how to tune it through, \textit{e.g.}, tailored chemical composition, which may open up alternative avenues for material design.\\
In summary, we investigated the role of thermal lattice vibrations on the band gap of \ch{CsPbBr3} as a representative, prototypical compound of HaP semiconductors.
The combination of DFT-based MC and MD calculations allowed for quantifying the impact of dynamic fluctuations beyond harmonic motions on the electronic structure, which was found to open the band gap by a large contribution of 450\,meV in the cubic phase at 425\,K.
We showed that the effect rationalizes the mild band-gap changes across temperature and phase-transitions that we recorded experimentally.
We conclude that pronounced anharmonic fluctuations play a leading role in the electronic structure of \ch{CsPbBr3}.
Other variants of HaPs can be expected to be influenced by it, too, since large-amplitude anharmonic motions of halide ions are a general property in these materials.\\
 \begin{acknowledgments}
\noindent We thank Alexey Chernikov (TU Dresden) for fruitful discussions.
Funding provided by the Alexander von Humboldt-Foundation in the framework of the Sofja Kovalevskaja Award, endowed by the German Federal Ministry of Education and Research, by
the Deutsche Forschungsgemeinschaft (DFG, German Research Foundation) via SPP2196 Priority Program (project-ID: 424709454) and via Germany's Excellence Strategy - EXC 2089/1-390776260, and by the Technical University of Munich - Institute for Advanced Study, funded by the German Excellence Initiative and the European Union Seventh Framework Programme under Grant Agreement No. 291763, are gratefully acknowledged.
The authors further acknowledge the Gauss Centre for Supercomputing e.V. for funding this project by providing computing time through the John von Neumann Institute for Computing on the GCS Supercomputer JUWELS at J\"ulich Supercomputing Centre.
O.Y. acknowledges funding from European Research Council (850041 — ANHARMONIC).
 \end{acknowledgments}

\appendix
\section{\label{app:a}Computational Methods}

\noindent \textcolor{Black}{All DFT calculations were performed with \texttt{VASP} \cite{kresse1996b}.
The core-valence interactions were treated using the projector-augmented wave method \cite{blochlProjectorAugmentedwaveMethod1994}, where semicore Pb-6$s$, Cs-5$p$ and Cs-6$s$ were treated as valence electrons.
We used the PBE functional \cite{perdew1996a} to describe exchange-correlation interactions.
For the MC method, applied to cubic $Pm\bar{3}m$ (space group 221) and orthorhombic $Pbnm$ (space group 62) phases of CsPbBr$_3$, we used dispersive corrections according to the Tkatchenko-Scheffler (TS) scheme \cite{tkatchenkoAccurateMolecularVan2009}.
For the MD calculations of both phases, we applied the TS scheme using an iterative Hirshfeld partitioning of the charge density \cite{buckoImprovedDensityDependent2013, buckoExtendingApplicabilityTkatchenkoScheffler2014}.
We applied different variants of dispersive corrections because it provided numerically stable calculations at all considered temperatures.
For structural relaxations, a plane-wave kinetic energy cutoff of 400~eV, an energy convergence threshold of $10^{-6}$~eV, and a $8\times8\times8$ $\Gamma$-centered $k$-grid ($6\times4\times6$) were used for the cubic (orthorhombic) phase.
The ionic and lattice degrees of freedom were optimized until the maximum residual forces on the atoms were below $5\times10^{-3}$~eV$\mathrm{\AA}^{-1}$.\\
DFT-MD simulations of CsPbBr$_3$ were performed for 160 atoms, \textit{i.e.}, a $4\times4\times2$ cubic and a $2\times2\times2$ orthorhombic supercell.
A Nosé-Hoover thermostat as implemented in \texttt{VASP}, with a time step of 8~fs, was used to control the temperature within the canonical (NVT) ensemble.
For these calculations, it was sufficient to apply a kinetic-energy cutoff of 300~eV, using a $1\times1\times2$ $k$-grid for the cubic and a single $k$-point for the orthorhombic phase.
The systems were equilibrated for 8~ps and the entire production run contains ($\sim$42\,ps), of which the first
16~ps were used to sample structures for band gap calculations.\\
A set of 100 distorted structures was obtained using the MC method as implemented in \texttt{VASP} \cite{karsaiElectronPhononCoupling2018} on the same supercells as were used in MD.
Specifically, phonon calculations within the harmonic approximation were performed in \texttt{VASP} to obtain the modes at the $\Gamma$-point.
Because each phonon mode is considered to be a harmonic oscillator, the probability distribution of displacements is a temperature-dependent Gaussian (see Ref.~\cite{SM}).
Samples of distorted structures are generated by summing over all the phonon eigenmodes with random, temperature-dependent amplitudes and phases.
Importantly, imaginary modes are ignored in this procedure.
Finally, we determine the band gap by taking the average over the DFT-calculated gap across the sample of structures.
The MC calculations were performed using a 400~eV kinetic-energy cutoff and an energy convergence threshold of $10^{-8}$~eV.
The integration in reciprocal space was carried out over a $3\times2\times3$ $k$-grid for the orthorhombic, and a $4\times4\times2$ grid for the cubic phase.\\
The fundamental band gap was calculated at different temperatures as a statistical average of 100 instantaneous configurations along the MD trajectory and MC-distorted structures.
The band gap was calculated using a kinetic-energy cutoff of 300~eV, an energy convergence threshold of $10^{-4}$~eV, and a $\Gamma$-centered $k$-grid of $2\times2\times2$ for orthorhombic, and $4\times4\times2$ for cubic CsPbBr$_3$.
We included the effect of SOC in these self-consistent DFT calculations.
}
\section{Experimental Methods}\label{app:b}

\noindent \textcolor{Black}
{CsPbBr$_3$ single crystals were grown using the slow vapor saturation of an antisolvent (VSA) method \cite{Rakita2016}.
3.832~g of CsBr (99.9\% trace metals basis, Sigma-Aldrich) and 6.608~g PbBr2 (98\%, Sigma-Aldrich) were dissolved in 40~mL dimethyl sulfoxide (DMSO, for HPLC, 99.7\%, Sigma-Aldrich).
The solution was heated on a hot plate that was set to 50$^\circ$ C till full dissolution.
The solution was then cooled to room temperature, and 40~mL of acetonitrile were added using a 5~mL automatic pipette (the addition of the acetonitrile lead to the formation of a light orange precipitate). \\
The mixture was then heated (50$^\circ$ C) and stirred for 24 hours, after which it was filtered using a 0.02 $\mathrm{\mu}$m filter.
The ratio of 1:4 solution:acetonitrile was used in the VSA crystallization systems (1~mL of filtered solution) inside a 4-mL vial, which was placed in an 18-mL vial with 4~mL of Acetonitrile.
After a week, the crystals were extracted and dried using blotting paper.
If the crystals were not used immediately, they were kept in the inner vial of the VSA system (within the solvent/antisolvent mixture) while capping it tightly.\\
CsPbBr$_3$ single crystals were taken into a glovebox under N$_2$ environment and placed in a high-temperature controller (Linkam, TS1000).
Measurements were conducted in a home-built back-scattering system \cite{asher2020anharmonic,sharma2020lattice}, using a white light source (250~W Quartz-Tungsten-Halogen lamp, Newport) set to 200~W.
The incident beam was directed into a microscope (Zeiss, USA), and focused on the sample through a 0.55 NA/50x objective (Zeiss, USA).
The back-reflected beam was collected by the objective, coupled to a fiber, and focused on a 1~m long spectrometer (FHR 1000, Horiba) dispersed by 150~gr/mm grating, achieving $\approx$0.8~meV spectral resolution, and detected by Si CCD (Horiba Inc., USA).
Reflectance measurements were first taken from the CsPbBr$_3$ surface, and consequently two additional reflectance measurements were taken -- from the Linkam's optical window (fused silica) and a background noise measurement (taken by blocking the white light beam close to the source).
Using the Linkam's optical window as a reference serves two objectives:
first, comparing the heated sample to a non-heated reference in order to not eliminate temperature effects of the sample.
Second, to measure at the same point on the crystal surface, achieved by changing only the $z$ axis focus and fixing the sample's $x$ and $y$ axes.
These three reflectance measurements -- sample, reference and background (abbreviated "S", "ref", and "BG", respectively) -- were the basic set of measurements taken at all temperatures, where the measurement parameters were kept fixed among the three.
Heating of the sample between measurements was carried out using a temperature ramp of 1$^\circ$ C/min with thermalization time of 5 minutes upon reaching the required temperature.
To ensure that no photo-damage is being done by the white light, a test was conducted by continuously exposing the crystal surface to the focused light for more than 7 minutes.
The test showed that there is no photo-damage of the crystal, both by the surface color and by comparing the reflectance spectra before and after exposure.\\
To calculate the reflectance contrast, which is the relative increase of the reflectivity compared to the reference, we used the following relation \cite{chernikov2015population,chernikov2014exciton}
\begin{equation*}
    R_c=\frac{R_S-R_{ref}}{R_{ref}-R_{BG}}.
\end{equation*}
Here, $R_c$ is the reflectance contrast, $R_S$, $R_{ref}$ and $R_{BG}$ are the reflectance from the sample surface, fused silica reference and background, respectively.\nocite{stoumpos2013,sajedi2022,heydHybridFunctionalsBased2003,krukauInfluenceExchangeScreening2006}
}
\bibliographystyle{apsrev4-1}
\bibliography{gap2023}
\end{document}